\documentclass[journal]{IEEEtran}
\usepackage{graphicx}
\usepackage{hyperref}
\usepackage{algorithm}
\usepackage{array}
\usepackage{multirow}

\ifCLASSINFOpdf
 
\else
  
\fi

\begin{document}
\title{Linear Nearest Neighbor Synthesis of Reversible Circuits by Graph Partitioning}
%
%
%

\author{Amlan Chakrabarti, Susmita Sur-Kolay, Senior Member, IEEE and Ayan Chaudhury

\thanks{Amlan Chakrabarti is with A.K.Choudhury School of I.T., University of Calcutta, 92, A.P.C.Road, 
Calcutta 700 009, India. e-mail: acakcs@caluniv.ac.in}
\thanks{Susmita Sur-Kolay is with Advanced Computing \& Microelectronics Unit, Indian Statistical Institute,
203, B.T.Road, Calcutta 700 108, India. e-mail: ssk@isical.ac.in }
\thanks{Ayan Chaudhury is with the Department
of Computer Science \& Engineering, University of Calcutta, 92, A.P.C. Road, 
Calcutta 700 009 , India. e-mail: ayanchaudhury.cs@gmail.com }}

\maketitle

\begin{abstract}
Linear Nearest Neighbor (LNN) synthesis in reversible circuits has emerged as an 
important issue in terms of technological implementation for quantum
computation. The objective is to obtain a LNN architecture with minimum gate
cost. As achieving optimal synthesis is a hard problem, heuristic methods have
been proposed in recent literature. In this work we present a graph
partitioning based approach for LNN synthesis with reduction in circuit cost.
In particular, the number of SWAP gates required to convert a given gate-level
quantum circuit to its equivalent LNN configuration is minimized. Our
algorithm determines the reordering of  indices of the qubit
line(s) for both single control and multiple controlled gates.
Experimental results for placing the target qubits of Multiple
Controlled Toffoli (MCT) library of benchmark circuits show a significant
reduction in gate count and quantum gate cost compared to those of related
research works.

\end{abstract}


\IEEEpeerreviewmaketitle

\section{Introduction}

\IEEEPARstart{Q}{uantum} computing is an emerging field of research in
which the rules of quantum physics are used to solve certain computing
problems more efficiently than any classical algorithm \cite{Sh98}.
A quantum circuit is employed to process quantum bits (qubit), where a qubit is the
unit of quantum information. It can be typically realized as a particular spin state
of an electron, or a certain polarization state of a photon. While the two
possible spin states of an electron are $up$ $(\uparrow)$ and $down$ $(\downarrow)$,
the two polarization states of a photon are $vertical$ ($\updownarrow$) and
$horizontal$ ($\leftrightarrow$ ). Such states are denoted by $|0 \rangle $ and $|1 \rangle $
respectively. A qubit can also represent any superposition state
$|\varphi \rangle $ = $\alpha|0 \rangle $ + $\beta|1 \rangle $ where $|\cdot \rangle $ is the standard Dirac notation \cite{F86}
for quantum states, and  $\alpha$  and $\beta$ are the complex amplitudes representing
the probabilities of state $|0 \rangle $ and $|1 \rangle $ respectively satisfying the condition $|\alpha|^2+|\beta|^2=1$.
Although a single qubit can theoretically be in infinite number of superposed states corresponding
to all possible pair of values for the amplitudes $\alpha$ and $\beta$, in reality equal amplitudes for
the two basis states lead to the observation that  $n$ qubits
can represent at most $2^n$ possible superposed states.
An \textit{n}-qubit quantum state can be represented as
\textit{$|\varphi \rangle = \sum_{i=0}^{2^n -1} \alpha_i |i \rangle  $}, where \textit{i} is written in binary
representation and $\sum_{i} |\alpha_i|^2 =1 $. This leads to the concept of a
quantum register \cite{NC02}
with \textit{n} qubits holding $2^n$ simultaneous values. Massive parallelism in
quantum computing can occur if an operation is performed on the contents of a
register, all possible values are operated on simultaneously. In
practice  however, the property of quantum de-coherence \cite{Pr73,Lo99} has to be tackled
in order to achieve this sort of parallelism.

\par

Quantum circuits or gate networks realizing Boolean functions are termed as
Quantum Boolean Circuits (QBCs). Formally,
a QBC is a quantum system of \textit{N} qubits specified by $|x_1 \rangle |x_2 \rangle  \ldots |x_N \rangle $, and a number of
reversible quantum gates. The quantum gates such as NOT, Controlled-Not (CNOT), SWAP, which implement
specific unitary operations manifest the logic operations. 

\par

Younnes and Miller \cite{Yo04} raised the
issue that only the interaction (coupling) between physically adjacent qubits is desired for practical
implementation of Quantum Boolean Circuits (QBCs). This is termed as the
{\it Linear Nearest-Neighbor (LNN)}
configuration. The requirement of nearest neighbor relationship between the control and the target qubits
is truly justified due to the limitation of the J-coupling force \cite{Gr05} required to perform multi-qubit logic
operations and this works effectively only between the adjacent qubits. In a
QBC, pairs of SWAP gates play a key role in bringing the control and the target
qubits of any quantum gate to adjacent lines. But this increases the gate cost.
Hence, the aim in LNN synthesis of minimizing the number of additional SWAP
gates has been addressed in this paper.

\par

The layout of the paper is as follows. The next section is a review of existing
synthesis approaches
including methods for cost reduction. In Section III, the proposed LNN synthesis
approach and related convention for multi-controlled gate decompositions are
discussed. In Section IV, the graph partitioning based LNN synthesis method is
presented. Experimental results appear next and concluding remarks in Section
VI.

\section{Related Works}

 One of the major challenges in hardware implementation by any technology is
to minimize power dissipation. Landauer \cite{La02}
showed that irreversible circuits must dissipate certain energy irrespective of
the technology used. Bennett \cite{Be02} in his work clearly established that the
most effective way of implementing reversible circuits to avoid this situation
is to employ quantum technology. Furthermore, using
quantum technology certain computational tasks can be performed
exponentially faster \cite{Sh94} than by the hardware of existing irreversible
technologies. 

\par

In the  implementation of quantum circuits, optimization of circuit levels as well gate count in a quantum boolean network needs
to be done. For a small circuit, this can be done easily, but for the larger ones, we need to find a proper algorithmic approach
for minimizing quantum cost in the circuit. Here, quantum cost means the cost in implementing a given quantum circuit using a suitable technology.

Synthesis of quantum boolean network is an emerging
research area and some fine works \cite{Mi02,Sh03,Ma04,Ma05} have been done on reversible circuit synthesis in recent years.
Local optimization methods have been proposed recently in \cite{Iw02,Vi05}.
Other models on reversible logic synthesis have also been proposed by many authors \cite{Mi06,Ya05,Hu06,Al10}. But 
from a practical viewpoint there are other
drawbacks with these approaches. Each gate must be realized in a Linear Nearest
Neighbor architecture. Wille \textit{et al.} \cite{Wi09} have discussed in detail about the cost metrics and the need for LNN
optimization. Hirata \textit{et al.} \cite{Hi09} discusses the schemes for permuting the qubits for achieving the LNN configuration.
 It carries out the optimization in the number of SWAP gates by carrying out the search for each gate individually, a sort 
of local ordering. It is not shown how this can lead to the optimization in the number of SWAP gates considering the whole circuit.
Related works can also be found in \cite{Pe11}. Most of these methods are built for Linear Nearest
Neighbor (LNN) architecture where only adjacent qubits can interact. In most of the articles, the authors aim at
minimizing nearest neighbor cost to achieve NNC optimality. Wille \textit{et al.} \cite{Wi09} showed that achieving NNC optimality by
introducing additional SWAP gates increases quantum cost. They aim optimality by decomposing
a TOFFOLI gate to basic gates, e.g, NOT, CNOT, Controlled-V and Controlled-V\textsuperscript{+} and then 
reordered the circuit lines by means of local and global reordering. This results in NNC optimality and reduced
quantum cost. Maslov \cite{Mo08} presents a theoretical study on quantum circuit placement in physical hardware by
means of a graph theoretic approach. A TOFFOLI network synthesis using template
matching has also been proposed in \cite{Ma05}.

\par

In this paper we propose a new approach for inserting pairs of SWAP gates, and
reordering the qubit lines based on
a graph partitioning approach. This gives better results compared to those by
the earlier ones including \cite{Wi09}.

\section{Linear Nearest Neighbor Synthesis}

Our work is based on reordering the qubit lines so that the interacting
qubits are adjacent to each other, i.e., the distance between target and control
lines are minimized (this is termed as Nearest Neighbor Cost (NNC) in \cite{Wi09}). 
Without loss of generality, it may be assumed that a given QBC is not in
nearest neighbor form. SWAP gates are inserted appropirately to
convert the QBC to a corresponding LNN architecture. It may be noted that for
1-qubit and 2-qubit gates, LNN architecture can be obtained without any
additional SWAP gates. Hence, the multi-qubit gates need special attention in
LNN synthesis to reduce the number of extra SWAP gates required,

In Fig. 1(a), a Toffoli or $C^k$NOT quantum gate is shown which is not in
the nearest neighbor architecture. Insertion of SWAP gates for LNN without
reordering the qubit lines is shown to be non-optimal in Fig. 1(b), whereas the
optimal synthesis after reordering as in Fig. 1(c) requires no SWAP gates.

\begin{figure}[h!]
  \centering
    \includegraphics[width=0.30\textwidth]{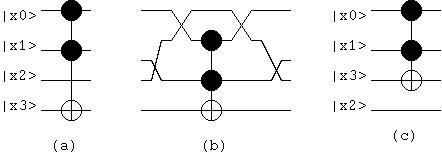}
    \caption{LNN synthesis of TOFFOLI (a) a single TOFFOLI gate, (b) introduction of SWAP gates resulting 4
             SWAPs. (c) is the reordered circuit resulting zero SWAP gate requirement}
\end{figure}

In this example of Fig. 1, we have considered a single gate where the reordering
can be done easily. The pertinent question is how to solve this for
large benchmark QBC circuits with many qubit lines. Furthermore,
an optimal solution with minimum NNC is the desideratum.

For ease of handling, we further sub-divide the LNN synthesis problem for QBCs
as one for NOT, CNOT and TOFFOLI (NCT) and another for multiple controlled TOFFOLI (MCT) gates. In the next two subsections we discuss
these two types of synthesis separately.

\subsection{LNN Synthesis for NCT gate library}

For the realization of quantum circuits in hardware technology we have to introduce SWAP gates or a
chain of basic quantum gates in the circuit to make it a LNN circuit. For introducing SWAP gates in
a TOFFOLI network to make a LNN architecture, an algorithm for counting the required SWAP gate pairs
is introduced here.

For a simple NOT gate, the number of SWAP pair required is zero. For a CNOT gate, the number of
SWAP pairs is simply the number of intermediate qubit lines between top and bottom control lines.
There may be two variations of a CNOT gate as shown in Fig. 2(a) and 2(b).

\begin{figure}[h!]
  \centering
    \includegraphics[width=0.35\textwidth]{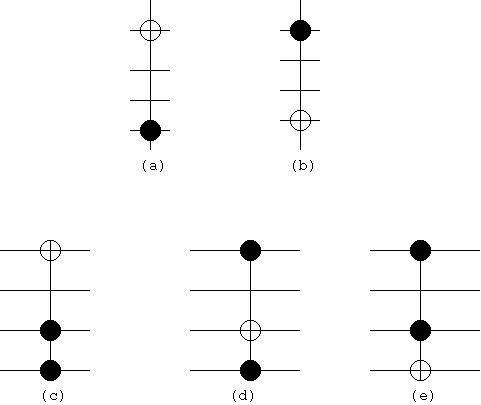}
    \caption{Variations of CNOT and TOFFOLI. (a), (b) are two types of CNOT. (c), (d), (e) are variations of TOFFOLI}
\end{figure}

For a TOFFOLI gate the scenario is somewhat complex. There may be three variations of a TOFFOLI gate
corresponding to the positions of its target qubit. All the variations of CNOT and TOFFOLI are shown
in Fig. 2(c), (d), (e).

\subsubsection{Rules for counting SWAP pairs}

We now formally present the rule for counting the additional SWAP gates needed to
convert a non-LNN gate to the corresponding LNN.
Let $ctrl_1$, $ctrl_2$, $target$ denote the two control qubit lines and the
target output qubit line of a TOFFOLI gate.
The input qubit lines have consecutive integer indices starting with $0$ for the
topmost qubit line. The control qubit line
having lower index value is denoted by $ctrl_1$ and the other control
qubit line  by $ctrl_2$.

\begin{center}
 \textbf{Rules for counting SWAP pairs}
\end{center}

\begin{tabbing}
{\bf Case 1:} \= $target < ctrl_1 < ctrl_2$\\
\> \= \textbf{if} $ctrl_1-target>1$, then required SWAP pair is \\
 \> \= $s_1+s_2$ where $s_1 = ctrl_1-target-1$ and \\
\> \= $s_2 = ctrl_2-target-2$ \\

\> \= \textbf{else if} $ctrl_2-ctrl_1>1$, then required SWAP pair \\
 \> \= is $s$ where $s = ctrl_2-ctrl_1-1$  \\

{\bf Case 2:} \= $ctrl_1 < target < ctrl_2$\\
\> \= \textbf{if} $target-ctrl_1>1$, then required SWAP pair is \\
 \> \= $s$ where $s = target-ctrl_1-1$, otherwise no \\
 \> \=  SWAP pair is needed \\

\> \= \textbf{if} $ctrl_2-target>1$, then required SWAP pair is \\
 \> \= $s$ where $s = ctrl_2-target-1$, otherwise no \\
 \> \=  SWAP pair is needed \\

{\bf Case 3:} \= $ctrl_1 < ctrl_2 < target$\\
\> \= \textbf{if} $target-ctrl_2>1$, then required SWAP pair is \\
 \> \= $s_1+s_2$ where $s_1 = target-ctrl2-1$ and \\
\> \= $s_2 = target-ctrl_1-2$ \\

\> \= \textbf{else if} $ctrl_2-ctrl_1>1$, then required SWAP pair \\
 \> \= is $s$ where $s = ctrl_2-ctrl_1-1$  \\

\end{tabbing}  

\subsection{LNN Synthesis for MCT gate library}

Another side of the problem is to decompose each MCT gate in the reversible circuit into an
equivalent NCT network. After then the nearest neighbor synthesis is to be done for the entire circuit.
Consider the MCT gate in Fig. 3. 

\begin{figure}[h!]
  \centering
    \includegraphics[width=0.25\textwidth]{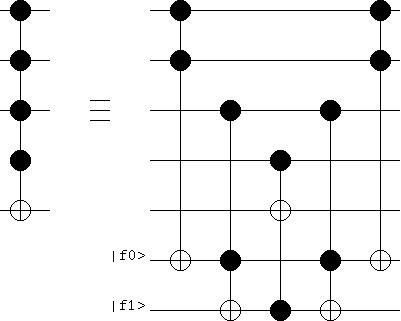}
    \caption{Decomposition of MCT gate: Replacement of a C\textsuperscript{4}NOT gate by equivalent
             TOFFOLI with two ancillary qubits $f_0$,$f_1$}
\end{figure}

Here, the original circuit is a
multiple controlled TOFFOLI. Using standard decomposition \cite{Pe11,Ac07} the gate is converted to equivalent TOFFOLI.
As a result, gate count and quantum cost has increased. Two extra ancillary qubits have come into the scenario
and the circuit is not in nearest neighbor condition.


As far as we are focussing the hardware implementation of quantum circuits, the decomposition of an arbitrary
circuit (Multiple controlled TOFFOLI) is necessary. Perkowski \textit{et al.} \cite{Pe11} states the necessity of decomposition and
introduction of ancillary qubits. Wille \textit{et al.} \cite{Wi09} showed some decomposition leading to increased
quantum cost by adding SWAP gates whenever a non-adjacent quantum gate appears.

\par

In our approach we have taken circuit files from \textit{Revlib} \cite{Rev} and written a C program
to parse it into an equivalent TOFFOLI network. It can be shown that for
decomposition of a single $C^k$NOT
gate, the number of TOFFOLI required is $2(k-2)+1$ and number of ancillary
qubits required is $k-2$ \cite{Ac07}. Using this
formula we have simply reconstructed the decomposed circuit. The resultant circuit contains only NOT, CNOT and TOFFOLI
gates.

\section{Reordering by Graph Partitioning}

In this section, we propose a graph partitioning based approach to get the ordering of qubit lines 
to achieve LNN architecture. Any non-LNN circuit can be converted to
LNN one by introducing additional SWAP gates. But the linear ordering of the qubit lines will make the difference
in number of SWAP gates required in making the LNN circuit. On a practical quantum computer, a number of
SWAP operations is necessary to emulate the behavior of a quantum circuit running on an ideal machine \cite{Ch11}.
Some works on TOFFOLI network synthesis can be found in \cite{Gr07,Ma07}. Some authors focused on logic synthesis by
permutation of qubit lines \cite{Dr09,Du03} also. Very recently LNN synthesis by means of line reordering has presented 
by Saeedi \textit{et al.} \cite{Sa11}.

However, in this section we propose a balanced graph partitioning approach through which reordering can be achieved with
less overhead  producing better result than the existing ones. Although graph partitioning problem is NP-complete, 
many algorithms exist that result a reasonably good partition. We first formulate the graph partitioning problem
mathematically  and then introduce the method how we have exploited the problem in NN synthesis.

\subsubsection{Qubit line adjacency graph}
We define an undirected weighted {\it qubit line adjacency graph}
$G=(V,E,w:E\rightarrow \cal N)$ for a given QBC $C$ where each vertex $v$
represents an input qubit line. There is an edge between two vertices $u$ and
$v$ if the corresponding qubit lines appear as the control and target qubit
lines of a single gate in $C$. Thus, for a NOT gate there is no
edge; for a CNOT gate there is an edge between its control and target qubit
lines; for a TOFFOLI$(ctrl_1, ctrl_2, target)$
gate there are two edges namely ($ctrl_1$,$target$) and ($ctrl_2, target$).
The weight $w(e)$ of an edge $e=(u,v)$ between the vertices $u$ and $v$ is the
number of times the pair of qubit lines corresponding to $u$ and $v$ appear
 as the  control and target qubit of any
quantum gate in the given QBC. It may noted that for a QBC with a single
output, this graph is connected. Fig. 4 illustrates a QBC of NCT gate
library and its  qubit line adjacency graph.

\begin{figure}[h!]
  \centering
    \includegraphics[width=0.30\textwidth]{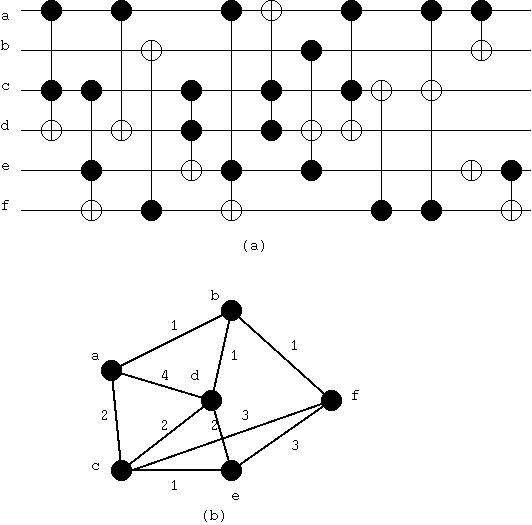}
    \caption{Graph formation: (a) is the TOFFOLI network having NOT, CNOT \& TOFFOLI gates, (b) is the
             weighted graph formed from this circuit}
\end{figure}

The problem of re-ordering the input qubit lines in a QBC $C$ is reduced to
finding a linear ordering $f: V \rightarrow \{1,2, \ldots, |V|\}$ of the
vertices of its qubit line adjacency graph $G$ such that $\Sigma_{e=(u,v)\in E}
|f(u) - f(v)|$ is minimum. This is known to be NP-complete \cite{GJ79}. There are
approximate algorithms for solving these \cite{FL07} which are non-trivial to
implement. Fortunately, it was also shown that the heuristic for getting
the linear ordering by listing the leaves of the binary tree
representing recursive balanced bi-partitioning of the graph, is good \cite{ENRS00}.
We therefore, adopt this line of approach to solve the problem of re-ordering
the qubit lines of a given QBC so that the NNC is minimal.

Various algorithms exist for solving graph partitioning problem in
polynomial time, having their
relative advantages and disadvantages. Many of these methods compute the eigenvector corresponding to the
second smallest eigenvalue \cite{Po92,Po90}. But these methods are expensive with respect to running time. Geometric partitioning
methods \cite{Rag95,Ga91} need greater execution time. Moreover, quality of partitions are also not up to the mark
due to randomized nature of partitioning of these approaches.

\subsubsection{Reordering by partitioning}

We have used \textit{pmetis} \cite{Pmet} graph partitioning tool for partitioning. This tool is developed
using multilevel partitioning algorithm (Coarsening, Partitioning and uncoarsening phases) and produce good
quality of partition with less execution time. The complexity of the algorithm used in \textit{pmetis} is
approximately $ O((n+m) * log(k))$ where $n$ is the number of nodes, $m$ is the number of edges and $k$ is the
number of partitions.

We have written a C program to parse the circuit description of a TOFFOLI network to the graph description format
required for \textit{pmetis} input. The output is a linear ordering of the partitions that are returned by \textit{pmetis}.
 But in most
of the cases the ordering returned by \textit{pmetis} there are fewer non-empty
partitions than the required number of partitions. Hence, more than one vertices may present in
a single partition. To resolve this problem, we have made a linear ordering of the partition order.  For example,
if the input qubit lines are 0 through 5 and the partitioning order is: $2, 1, 1, 3, 2, 4$, we will reorder these as:
$2, 0, 1, 4, 3, 5$.

\begin{figure}[h!]
  \centering
    \includegraphics[width=0.45\textwidth]{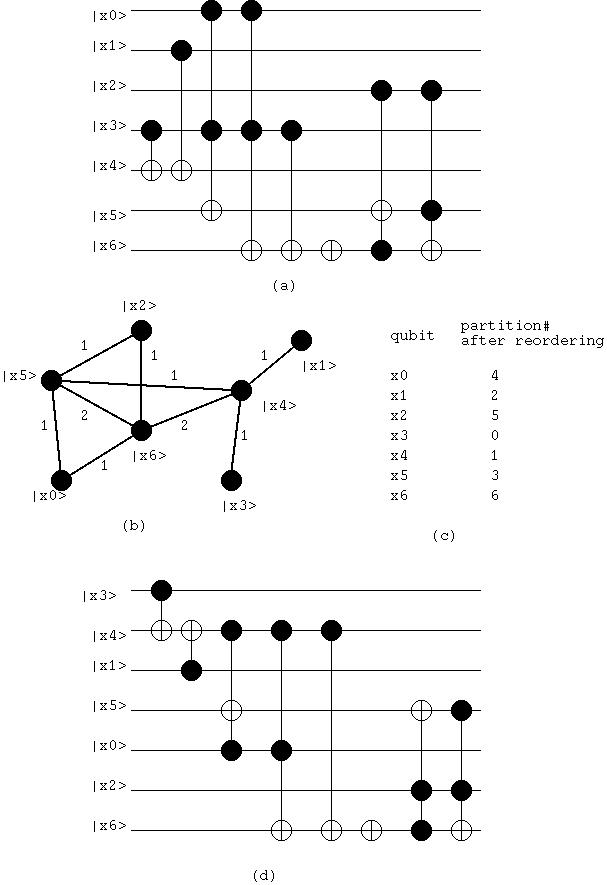}
    \caption{Synthesis of the circuit \textit{4mod5-bdd\_287}. (a) the original circuit which needs 15 SWAP pairs for LNN
             architecture, (b) the graph formed from reordered circuit in (d), (c) the ordering of qubit lines
             after partitioning, (d) is the reconstructed circuit from new ordering, with only 10 SWAP
             pairs for LNN architecture}
\end{figure}

From the reordered qubit lines we have again reconstructed the NCT network. Then the required  number of
SWAP pair is calculated in both cases using the rule presented in the previous section. Figure 5 demonstrates
the synthesis of the circuit \textit{4mod5-bdd\_287} from \textit{RevLib}.
The original circuit is represented in Fig.5(a), which is not in NN architecture.
The number of SWAP gates required to make this circuit LNN is 15. Fig.5(b) is the graph formed from reordered
circuit in (d) and 5(c) are the partition numbers returned by \textit{pmetis} after the partitioning.
Fig.5(d) is the reconstructed circuit after the reordering, which clearly shows that number of SWAP gate requirement
has come down to $10$ to make it an LNN circuit.

\section{Experimental Results}

The experimental results are shown
in Table \MakeUppercase{\romannumeral 1}. The first column is the list of benchmark circuits used in our
experiment, next three columns are the number of qubit lines (N), gate count (GC) and quantum cost (QC) in the original circuit.
SWAP cost and total quantum cost (QC) of the decomposed circuit before ordering are shown in the next two columns, 
followed by SWAP cost and QC after ordering. Then the percentage
reduction of quantum cost after reordering the qubit lines are shown. Next column denotes the results reported by
Wille \textit{et al.} \cite{Wi09}. The percentage decrease in quantum cost over \cite{Wi09} is shown in the last column.

Regarding the quantum cost of various gates, cost of NOT, CNOT and SWAP gates are taken as $1, 1$ and $3$ 
respectively \cite{Wi09}. When two qubits of a particular gate are adjacent, then there is no need to introduce SWAP between them,
when two qubit lines are adjacent then only the notion of introducing SWAP gates comes into
the scenario. We have
taken 54 benchmark circuits from \textit{RevLib}, out of which the results for 20 circuits are also reported in \cite{Wi09}.
It can be seen that using our method we can reduce the number of SWAP gates and therefore the total quantum
cost by 17.5\%  in average compared to \cite{Wi09}, except for a few of the cases in which
ordering of the qubit lines have worsened the situation. The execution times of the algorithm are also negligible.
 The percentage decrease in quantum cost due to the reordering of qubit lines, considering all the circuits as given
 in Table I is about 47\%. \\ 

\begin{table*}
\caption{Comparison of Quantum cost for proposed LNN synthesis method vs. \cite{Wi09} on RevLib Benchmarks} \label{table 1}
\begin{center}
\begin{tabular}{|p{2.5cm}||p{0.7cm}|p{0.7cm}|p{0.7cm}||p{1cm}|p{1cm}||p{1cm}|p{1cm}||p{1.7cm}||p{1.3cm}||p{1.5cm}|} \hline
& \multicolumn{3}{|p{2.4cm}||}{\textsc{Original Circuit}} & \multicolumn{2}{|p{2.4cm}||}{\textsc{Before ordering}} & \multicolumn{2}{|p{2.2cm}||}{\textsc{After ordering}}  & \multicolumn{1}{|c||}{} & \multicolumn{1}{|c||}{} & \\
\textsc{Benchmark Circuit} & \textsc{N} & \textsc{GC} & \textsc{QC} & \textsc{SWAP cost} & \textsc{Total QC} & \textsc{SWAP cost} & \textsc{Total QC} & \% \textsc{decrease in QC due to reordering} & \textsc{QC reported by \cite{Wi09}} & \% \textsc{decrease in QC over \cite{Wi09}} \\
\hline
\hline
3\_17\_13 & 3 & 6 & 14 & 6 & 20 & 0 & \bf{14} & 30 & 28 & 50 \\
\hline
4\_49\_17 & 4 & 12 & 16  & 66 & 98 & 18 & \bf{50} & 48.9 & 98 & 48.9 \\
\hline
4gt4-v0\_80 & 5 & 5 & 37  & 120 & 153 & 48 & \bf{81} & 47 & 138 & 41.3 \\
\hline
4gt5\_75 & 5 & 5 & 21 & 78 & 101 & 36 & \bf{59} & 41.5 & 79 & 25.3 \\
\hline
4gt12-v1\_89 & 5 & 5 & 45  & 114 & \bf{157} & 126 & 169 & -7.64 & 168 & 6.5 \\
\hline
4gt13-v1\_93 & 5 & 4 & 16  & 72 & 90 & 6 & \bf{24} & 73.3 & 53 & 54.7 \\
\hline
4gt-10v1\_81 & 5 & 6 & 34  & 120 & 158 & 96 & \bf{134} & 15.2 & 147 & 8.8 \\
\hline
4mod5-bdd\_287 & 7 & 8 & 24  & 90 & 114 & 72 & \bf{96} & 15.7 & - & -\\
\hline
4mod5-v1\_23 & 5 & 8 & 24  & 84 & 108 & 42 & \bf{66} & 38.8 & 78 & 15.3 \\
\hline
5xp1\_194 & 17 & 85 & 1430  & 28194 & 29523 & 7164 & \bf{8493} & 71.2 & - & -\\
\hline
9symml\_195 & 10 & 129 & 14193  & 34458 & 38303 & 19764 & \bf{23609} & 38.3 & - & -\\
\hline
add6\_196 & 19 & 229 & 6455  & 122910 & 128831 & 45606 & \bf{51527} & 60 & - & -\\
\hline
adr4\_197 & 13 & 55 & 727  & 9774 & 10489 & 4986 & \bf{5701} & 45.6 & - & -\\
\hline
aj-e11\_165 & 4 & 13 & 45  & 84 & \bf{131} & 90 & 137 & -4.5 & 181 & 27.6\\
\hline
alu1\_198 & 20 & 32 & 228  & 6498 & 6756 & 1944 & \bf{2202} & 67.4 & - & -\\
\hline
alu2\_199 & 16 & 157 & 5654 & 70716 & 74991 & 21240 & \bf{25515} & 65.9 & - & -\\
\hline
alu3\_200 & 18 & 94 & 2632  & 45954 & 48290 & 21000 & \bf{23336} & 51.6 & - & -\\
\hline
alu4\_201 & 22 & 1063 & 55388  & 1059834 & 1106423 & 541380 & \bf{587969} & 46.8 & - & -\\
\hline
alu-bdd\_288 & 7 & 9 & 29  & 144 & 173 & 102 & \bf{131} & 24.2 & - & -\\
\hline
apla\_203 & 22 & 80 & 3438  & 77742 & 80828 & 23394 & \bf{26480} & 67.2 & - & -\\
\hline
apex4\_202 & 28 & 5376 & 237963  & 7659894 & 7875016 & 1922358 & \bf{2137480} & 72.8 & - & -\\
\hline
bw\_291 & 87 & 307 & 943  & 79326 & 80269 & 46158 & \bf{47101} & 41.3 & - & -\\
\hline
c7552\_205 & 21 & 80 & 1728  & 50418 & 52102 & 10746 & \bf{12430} & 76.1 & - & -\\
\hline
clip\_206 & 14 & 174 & 6731  & 72792 & 77764 & 29292 & \bf{34264} & 55.9 & - & -\\
\hline
cm42a\_207 & 14 & 35 & 377  & 7236 & 7617 & 1590 & \bf{1971} & 74.7 & - & -\\
\hline
cm85a\_209 & 14 & 69 & 2252  & 26958 & 28995 & 11562 & \bf{13599} & 54 & - & -\\
\hline
cm150a\_210 & 22 & 53 & 1096  & 8472 & 9467 & 5124 & \bf{6119} & 35.3 & - & -\\
\hline
cm151a\_211 & 28 & 33 & 888  & 21216 & 22027 & 5016 & \bf{5827} & 73.5 & - & -\\
\hline
cm152a\_212 & 12 & 16 & 252  & 1566 & 1816 & 1416 & \bf{1666} & 8.2 & - & -\\
\hline
cm163a\_213 & 29 & 39 & 756  & 22800 & 23499 & 6780 & \bf{7479} & 68.1 & - & -\\
\hline
cmb\_214 & 20 & 18 & 910  & 7800 & 8234 & 2748 & \bf{3182} & 61.3 & - & -\\
\hline
co14\_215 & 15 & 30 & 3488  & 24438 & 26064 & 8466 & \bf{10092} & 61.2 & - & -\\
\hline
cu\_219 & 25 & 40 & 1148  & 30234 & 31262 & 9402 & \bf{10430} & 66.6 & - & -\\
\hline
cycle10\_2\_110 & 12 & 19 & 1202  & 9690 & 10417 & 2808 & \bf{3535} & 66 & 8046 & 56\\
\hline
dc1\_220 & 11 & 39 & 416  & 5994 & 6419 & 1482 & \bf{1907} & 70.2 & - & -\\
\hline
dc2\_222 & 15 & 75 & 1886  & 32052 & 33809 & 8634 & \bf{10391} & 69.2 & - & -\\
\hline
decod24-v3\_46 & 4 & 9 & 9  & 54 & 63 & 12 & \bf{21} & 66.6 & \bf{21} & 0 \\
\hline
decod\_217 & 21 & 80 & 1728  & 50418 & 52102 & 10746 & \bf{12430} & 76.1 & - & -\\
\hline
dist\_223 & 13 & 185 & 7601  & 76164 & 81591 & 24828 & \bf{30255} & 62.9 & - & -\\
\hline
f51m\_233 & 22 & 663 & 37400  & 612030 & 639493 & 281508 & \bf{308971} & 51.6 & - & -\\
\hline
ham15\_108 & 15 & 70 & 453  & 3312 & 3764 & 2418 & 2870 & 23.7 & \bf{2588} & -10.8 \\
\hline
hwb4\_52 & 4 & 11 & 23  & 42 & \bf{65} & 48 & 71 & -9.2 & \bf{65} & 0\\
\hline
hwb5\_55 & 5 & 24 & 104  & 378 & 492 & 276 & 390 & 20.7 & \bf{337} & -15.7 \\
\hline
inc\_237 & 16 & 93 & 2140  & 40410 & 42407 & 11772 & \bf{13769} & 67.5 & - & -\\
\hline
mod5adder\_128 & 6 & 15 & 83  & 522 & 613 & 150 & \bf{241} & 60.6 & 330 & 26.9\\
\hline
mod8-10\_177 & 5 & 14 & 94  & 330 & 418 & 234 & \bf{322} & 22.9 & 363 & 11.2\\
\hline
plus127mod8192\_162 & 13 & 910 & 73357  & 508134 & 551588 & 403938 & \bf{447392} & 18.8 & 503516 & 11.1\\
\hline
plus63mod4096\_163 & 12 & 429 & 32539  & 193446 & 211559 & 159258 & \bf{177371} & 16.1 & 210400 & 15.6\\
\hline
plus63mod8192\_164 & 13 & 492 & 45025  & 267480 & 290798 & 203772 & \bf{227090} & 21.9 & 279016 & 18.6\\
\hline
rd53\_135 & 7 & 16 & 77  & 558 & 636 & 348 & 426 & 33 & \bf{303} & -40.5 \\
\hline
rd84\_313 & 34 & 104 & 304  & 6780 & 7084 & 4548 & \bf{4852} & 31.5 & - & -\\
\hline
sqn\_258 & 10 & 76 & 2122  & 15258 & 16784 & 5562 & \bf{7088} & 57.7 & - & -\\
\hline
sym9\_317 & 27 & 62 & 206  & 3870 & 4076 & 1686 & \bf{1892} & 53.5 & - & -\\
\hline
z4ml\_269 & 11 & 48 & 642  & 7386 & 8018 & 3600 & \bf{4232} & 47.2 & - & -\\
\hline
\end{tabular}

\begin{tabular}{|p{11.9cm}||p{3.5cm}||p{1.5cm}|} \hline
\hline
\textsc{Average cost reduction \%}  & \bf{46.6} & \bf{17.5} \\
\hline

\end{tabular}

\end{center}
\end{table*}

\section{Conclusion}
In this work, we have focussed on linear nearest neighbor architecture where SWAP gates are introduced between the qubit lines
whenever it is not in NN architecture. The method proposed reduces the SWAP gate count significantly for the RevLib benchmark circuits. It has also led to 
an appreciable decrease in the overall quantum cost for most of the benchmark circuits compared to the related works, and by 47\% on an average. This method for converting a QBC
with CNOT, C\textsuperscript{2}NOT and even MCT gates into their nearest neighbor equivalent form can be utilized for the development of
low-level circuit synthesis automation tools in the quantum computing domain.


\begin{thebibliography}{1}


\bibitem{Sh98}
P. W. Shor, ``\emph{Quantum Computing}'', Documenta Mathematica-Extra Volume, ICM, pp. 1-1000, 1998.

\bibitem{F86}
R. P. Feynman, ``\emph{Quantum Mechanical Computers,}'', Foundations of Physics,
Vol.16, pp.507-531, 1986.

\bibitem{NC02}
M. A. Nielsen and Isaac L. Chuang, 
``\emph{Quantum Computation and Quantum Information}'', Cambridge
University Press, 2002.


\bibitem{Pr73}
J. Preskil, 
``\emph{Logical reversibility of computation,}'', I.B.M. J. Res. Dev., Vol.17, pp. 525-532, 1973.

\bibitem{Lo99}
H. K. Lo, S. Popescu and T. Spiller, 
``\emph{Introduction to quantum computation and information,}'', Singapore : World Scientific Publ., 1999.

\bibitem{Yo04}
A. Younes and J. Miller, 
``\emph{Representation of Boolean Quantum Circuits as Reed Muller Expressions,}'',
International Journal of Electronics, Volume 91,  No. 7, pp. 431-444, July 2004.


\bibitem{Gr05}
A. Grigorenko and D. V. Khveshchenko, 
``\emph{Single Step Implementation of Universal Quantum Gates,}'',
Physical Review Letters, 95.110501, 2005.

\bibitem{La02}
R. Landauer, 
``\emph{Irreversibility and heat generation in the computing process,}'',
IBM Journal of Research and Development, 5:183-191, July 1961.

\bibitem{Be02}
C. Bennett., 
``\emph{Logical reversibility of computation,}'', I.B.M. J. Res. Dev., Vol.17, pp. 525-532, 1973.


\bibitem{Sh94}
Shor, P. (1994) ``\emph{Algorithms for quantum computation: Discrete logarithms and factoring,}'', Proceedings of the 35th Annual
IEEE Symposium on Foundations of Computer Science, pp. 124-134

\bibitem{Mi02}
D. M. Miller, 
``\emph{Spectral and two-place decomposition techniques in reversible logic,}'',
In Midwest Symposium on Circuits and Systems, August 2002.

\bibitem{Sh03}
V. V. Shende, A. K. Prasad, I. L. Markov, and J. P. Hayes, 
``\emph{Synthesis of reversible logic circuits,}'',
IEEE Transactions on CAD, 22(6):723-729, June 2003.


\bibitem{Ma04}
D. Maslov and G. W. Dueck, 
``\emph{Reversible cascades with minimal garbage,}'',
IEEE Transactions on CAD, 23(11):1497-1509, November 2004.

\bibitem{Ma05}
D. Maslov, G. W. Dueck, and D. M. Miller, 
``\emph{Toffoli network synthesis with templates,}'',
IEEE Transactions on CAD, 24(6):807-817, June 2005.

\bibitem{Iw02}
K. Iwama, Y. Kambayashi, and S. Yamashita, 
``\emph{Transformation rules for designing CNOT-based quantum circuits,}'',
In Design Automation Conference, pages 419-424, New Orleans, Louisiana, USA, June 2002.


\bibitem{Vi05}
G. F. Viamontes, I. L. Markov, and J. P. Hayes, 
``\emph{Graph based simulation of quantum computation in the state vector and density matrix representation,}'',
Quantum Information and Computation, Vol. 5, No. 2, pp. 113-130, 2005.

\bibitem{Mi06}
D. M. Miller, D. Maslov, and G. W. Dueck, 
``\emph{Synthesis of quantum multiple valued circuits,}'',
Journal of Multiple-Valued Logic and Soft Computing, vol. 12, no. 5-6, pp. 431-450, 2006.

\bibitem{Ya05}
G. Yang, W. Hung, X. Song, and M. Perkowski, 
``\emph{Majority-based reversible logic gates,}'',
Theoretical Computer Science, vol. 334, no. 1-3, 2005.


\bibitem{Hu06}
W. N. N. Hung, X. Song, G. Yang, J. Yang, and M. Perkowski, 
``\emph{Optimal synthesis of multiple output boolean functions using a set of quantum gates
by symbolic reachability analysis,}'',
IEEE Transactions on CAD, vol. 25, no. 9, pp. 1652-1663, 2006.

\bibitem{Al10}
W. N. Alhagi, M. Hawash, and M. Perkowski, 
``\emph{Synthesis of Reversible Circuits for Large Reversible Functions,}'',
Facta Universitais, Series: Electronics and Energetics,
vol. 24, no. 3, pp. 273-289, December 2010.

\bibitem{Wi09}
R. Wille, M. Saeedi and R. Drechsler, 
``\emph{Synthesis of Reversible Functions Beyond Gate Count and Quantum Cost,}'',
International Workshop on Logic Synthesis (IWLS), USA, 2009.

\bibitem{Hi09}
Y. Hirata, M. Nakanishi, and S. Yamashita, 
``\emph{An efficient method to convert arbitrary quantum circuits to ones on a
linear nearest neighbor architecture,}'',
3rd International conference on Quantum, Nano, and Micro Technologies, 2009.


\bibitem{Pe11}
M. Perkowski, M. Lukac, D. Shah, and M. Kameyama, 
``\emph{Synthesis of quantum circuits in Linear Nearest Neighbor model using Positive Davio Lattices,}'',
Facta Universitais, Series: Electronics and Energetics, vol. 24, no. 1, pp. 71-87, April 2011.

\bibitem{Mo08}
D. Maslov, S.M. Falconer, and M. Mosca, 
``\emph{Quantum Circuit Placement,}'',
IEEE Transactions on CAD, Volume. 27, No. 4, 752 - 763, April 2008.

\bibitem{Ac07}
A. Chakrabarti and S. Sur-Kolay, 
``\emph{Nearest Neighbour based Synthesis of Quantum Boolean Circuits,}'',
Engineering Letters, vol. 15, no. 2, 2007.

\bibitem{Rev}
http://www.revlib.org/

\bibitem{Ch11}
B.S. Choi and R. Van Meter, 
``\emph{Effects of Interaction Distance on Quantum Addition Circuits,}'',
ACM Journal on Emerging Technologies in Computing Systems, Volume 7, Issue 3, August 2011.

\bibitem{Gr07}
D. Große, X. Chen, G.W. Dueck, R. Drechsler, 
``\emph{Exact SAT-based Toffoli Network Synthesis,}'',
Great Lakes Symposium on VLSI (GLSVLSI), pp. 96-101, Stresa, 2007.

\bibitem{Ma07}
D. Maslov, D. M. Miller, G. W. Dueck,
``\emph{Techniques for the Synthesis of Reversible Toffoli Networks,}'',
ACM Transactions on Design Automation of Electronic Systems, Volume 12, Issue 4, September 2007 .

\bibitem{Dr09}
R. Wille, D. Große, G.W. Dueck, R. Drechsler, 
``\emph{Reversible Logic Synthesis with Output Permutation,}'',
22nd International Conference on VLSI Design, pp. 189-194, New Delhi, 2009.

\bibitem{Du03}
D.M. Miller, D. Maslov, G. W. Dueck, 
``\emph{A Transformation Based Algorithm for Reversible Logic Synthesis,}'',
Proceedings of the 40th annual Design Automation Conference(DAC), June 2-6, 2003, Anaheim, California, USA.

\bibitem{Sa11}
M. Saeedi, R. Wille, and R. Drechsler,
``\emph{Synthesis of quantum circuits for linear nearest neighbor architectures,}'',
Quantum Information Processing, Volume 10 Issue 3, June 2011.

\bibitem{GJ79}
M. R. Garey And D. S. Johnson 1979, 
``\emph{Computers and Intractability: A Guide to Theory of NP Completeness,}'', W. H. Freeman and Co., San Francisco, CA.

\bibitem{FL07} U. Feige  and J. R. Lee, ``\emph{An improved approximation ratio for the minimum linear arrangement problem,}'',
Inform. Process. Lett. 101, 1, 26-29, 2007.

\bibitem{ENRS00} G. Even, Joseph S. Naor, S. Rao and B. Schieber, 
``\emph{Divide-and-conquer approximation algorithms via spreading metrics,}'' Journal of the
ACM (JACM), vol. 47, no. 4, pp. 585-616, July 2000.


\bibitem{Po92}
A. Pothen, H. D. Simon, L. Wang, and S. T. Bernard,
``\emph{Towards a fast implementation of spectral nested dissection,}'',
In Supercomputing '92 Proceedings, pages 42-51, 1992.

\bibitem{Po90}
A. Pothen, H. D. Simon, and K. P. Liou,
``\emph{Partitioning sparse matrices with eigenvectors of graphs,}'', 
SIAM Journal of Matrix Analysis and Applications, 11(3):430-452, 1990.

\bibitem{Rag95}
M. T. Heath and P. Raghavan,
``\emph{A Cartesian parallel nested dissection algorithm,}'', 
SIAM Journal of Matrix Analysis and Applications, 16(1):235-253, 1995.

\bibitem{Ga91}
G. L. Miller, S. H. Teng, and S. A. Vavasis,
``\emph{A unified geometric approach to graph separators,}'',
In Proceedings of 31st Annual Symposium on Foundations of Computer Science, pages 538-547, 1991.

\bibitem{Pmet}
http://glaros.dtc.umn.edu/gkhome/metis/metis/overview/.



\end{thebibliography}

\end{document}